\documentclass[a4paper,11pt]{article}

\usepackage{amsmath}
\usepackage{slashed}

\usepackage{amsmath,amsthm,amsfonts,amssymb, graphicx,mathabx,epsfig}
\usepackage{bbm}
\usepackage{authblk}
\usepackage{bm}
\usepackage{graphicx,color}
\usepackage{footmisc}
\usepackage{epstopdf}
\usepackage{dsfont}
\usepackage{pgfplots}
\usepackage{subfig}
\usetikzlibrary{fit}
\usepackage{slashed}

\usepackage{graphicx}
\usepackage{dcolumn}
\usepackage{bm}
\usepackage{epstopdf}

\newcount\driver
\newcount\bozza

scaled\magstep1                  
scaled\magstep1 \font\msytw=msbm10 scaled\magstep1
 
scaled\magstep1                 \font\indbf=cmbx10 scaled\magstep2

scaled \magstep2

{\count255=\time\divide\count255 by 60
\xdef\hourmin{\number\count255}
        \multiply\count255 by-60\advance\count255 by\time
   \xdef\hourmin{\hourmin:\ifnum\count255<10 0\fi\the\count255}}

\let\a=\alpha \let\b=\beta    \let\g=\gamma     \let\d=\delta     \let\e=\varepsilon
\let\z=\zeta  \let\h=\eta     \let\th=\vartheta      \let\l=\lambda
\let\m=\mu    \let\n=\nu      \let\x=\xi                \let\r=\rho
\let\s=\sigma \let\t=\tau            
\let\ps=\psi   \let\o=\omega     
 \let\D=\Delta       \let\L=\Lambda

\def\EE{{\cal E}}\def\VV{{\cal V}}
\def\HH{{\cal H}}
\def\TT{{\cal T}}\def\BB{{\cal B}}
\def\RR{{\cal R}}\def\LL{{\cal L}}

\def\pp{{\bf p}}\def\xx{{\bf x}}
  
\def\yy{{\bf y}}\def\nn{{\bf n}}

       \def\oo{{\underline \omega}}
\def\ee{{\underline \varepsilon}}

          \def\BBB{\hbox{\euftw B}}
\def\RRR{\hbox{\msytw R}}

        \def\ZZZ{\hbox{\msytw Z}}
        
\def\TTT{\hbox{\msytw T}}          
        \def\EE{\hbox{\msytw E}}

\let\==\equiv

\let\io=\infty
\let\0=\noindent

\def\*{{\hfill\break\null\hfill\break}}

\def\tilde#1{{\widetilde #1}}

\def\la{{\langle}}
\def\ra{{\rangle}}

\def\tende#1{\,\vtop{\ialign{##\crcr\rightarrowfill\crcr
             \noalign{\kern-1pt\nointerlineskip}
             \hskip3.pt${\scriptstyle #1}$\hskip3.pt\crcr}}\,}
\def\otto{\,{\kern-1.truept\leftarrow\kern-5.truept\to\kern-1.truept}\,}

\def\Tr{\rm Tr}

\def\wh#1{\widehat{#1}}
\def\hat#1{\wh{#1}}
\def\sqt[#1]#2{\root #1\of {#2}}

\def\bp{{\bar \ps}}

\def\EE{{\cal E}}\def\VV{{\cal V}}
\def\HH{{\cal H}}
\def\TT{{\cal T}}\def\BB{{\cal B}}
\def\RR{{\cal R}}\def\LL{{\cal L}}

\def\T#1{{#1_{\kern-3pt\lower7pt\hbox{$\widetilde{}$}}\kern3pt}}
\def\VVV#1{{\underline #1}_{\kern-3pt
\lower7pt\hbox{$\widetilde{}$}}\kern3pt\,}
\def\W#1{#1_{\kern-3pt\lower7.5pt\hbox{$\widetilde{}$}}\kern2pt\,}

\def\indica{\leaders \hbox to 0.5cm{\hss.\hss}\hfill}
\def\guida{\leaders\hbox to 1em{\hss.\hss}\hfill}
\mathchardef\oo= "0521

\def\pp{{\bf p}}\def\xx{{\bf x}}
\def\yy{{\bf y}}\def\nn{{\bf n}}

\def\oo{{\underline \omega}}

\def\qed{\raise1pt\hbox{\vrule height5pt width5pt depth0pt}}
 
  \def\bp{{\bar p}} 

\def\indic{\hbox{\raise-2pt \hbox{\indbf 1}}}

\def\RRR{\hbox{\msytw R}}

 \def\ZZZ{\hbox{\msytw Z}}
 
\def\TTT{\hbox{\msytw T}}

\def\ins#1#2#3{\vbox to0pt{\kern-#2 \hbox{\kern#1 #3}\vss}\nointerlineskip}

\newdimen\xshift \newdimen\xwidth \newdimen\yshift

\def\insertplot#1#2#3#4#5#6{%
\xwidth=#1pt \xshift=\hsize \advance\xshift by-\xwidth \divide\xshift by 2%
\begin{figure}[ht]
\vspace{#2pt} \hspace{\xshift}
\begin{minipage}{#1pt}
#3 \ifnum\driver=1 \griglia=#6
\ifnum\griglia=1 \openout13=griglia.ps \write13{gsave .2
setlinewidth} \write13{0 10 #1 {dup 0 moveto #2 lineto } for}
\write13{0 10 #2 {dup 0 exch moveto #1 exch lineto } for}
\write13{stroke} \write13{.5 setlinewidth} \write13{0 50 #1 {dup 0
moveto #2 lineto } for} \write13{0 50 #2 {dup 0 exch moveto #1
exch lineto } for} \write13{stroke grestore} \closeout13
\includegraphics{griglia.ps} \fi
\includegraphics{#4.ps}\fi%
\ifnum\driver=2 \fi
\end{minipage}
\caption{#5}
\end{figure}
}

\newdimen\shift \shift=-1.5truecm
\def\lb#1{%
\ifnum\bozza=1
\label{#1}\rlap{\hbox{\hskip\shift$\scriptstyle#1$}}
\else\label{#1} \fi}

\def\be{\begin{equation}}
\def\ee{\end{equation}}
\def\bea{\begin{eqnarray}}\def\eea{\end{eqnarray}}
\def\bean{\begin{eqnarray*}}\def\eean{\end{eqnarray*}}
\def\bfr{\begin{flushright}}\def\efr{\end{flushright}}
\def\bc{\begin{center}}\def\ec{\end{center}}
\def\bal{\begin{align}}\def\eal{\end{align}}
\def\ba#1{\begin{array}{#1}} \def\ea{\end{array}}
\def\bd{\begin{description}}\def\ed{\end{description}}
\def\bv{\begin{verbatim}}\def\ev{\end{verbatim}}
\def\nn{\nonumber}
\def\Halmos{\hfill\vrule height10pt width4pt depth2pt \par\hbox to \hsize{}}
\def\pref#1{(\ref{#1})}

\def\ins#1#2#3{\vbox to0pt{\kern-#2 \hbox{\kern#1 #3}\vss}\nointerlineskip}

\newdimen\xshift \newdimen\xwidth \newdimen\yshift
\newcount\griglia

\def\insertplot#1#2#3#4#5#6{%
\xwidth=#1pt \xshift=\hsize \advance\xshift by-\xwidth \divide\xshift by 2%
\begin{figure}[ht]
\vspace{#2pt} \hspace{\xshift}
\begin{minipage}{#1pt}
#3 \ifnum\driver=1 \griglia=#6
\ifnum\griglia=1 \openout13=griglia.ps \write13{gsave .2
setlinewidth} \write13{0 10 #1 {dup 0 moveto #2 lineto } for}
\write13{0 10 #2 {dup 0 exch moveto #1 exch lineto } for}
\write13{stroke} \write13{.5 setlinewidth} \write13{0 50 #1 {dup 0
moveto #2 lineto } for} \write13{0 50 #2 {dup 0 exch moveto #1
exch lineto } for} \write13{stroke grestore} \closeout13
\includegraphics{griglia.ps} \fi
\includegraphics{#4.ps}\fi%
\ifnum\driver=2 \fi
\end{minipage}
\caption{#5}
\end{figure}
}

\newdimen\shift \shift=-1.5truecm
\def\lb#1{%
\label{#1}\rlap{\hbox{\hskip\shift$\scriptstyle#1$}}
\else\label{#1} \fi}

\def\be{\begin{equation}}
\def\ee{\end{equation}}
\def\bea{\begin{eqnarray}}\def\eea{\end{eqnarray}}
\def\bean{\begin{eqnarray*}}\def\eean{\end{eqnarray*}}
\def\bfr{\begin{flushright}}\def\efr{\end{flushright}}
\def\bc{\begin{center}}\def\ec{\end{center}}
\def\bal{\begin{align}}\def\eal{\end{align}}
\def\ba#1{\begin{array}{#1}} \def\ea{\end{array}}
\def\bd{\begin{description}}\def\ed{\end{description}}
\def\bv{\begin{verbatim}}\def\ev{\end{verbatim}}
\def\nn{\nonumber}
\def\Halmos{\hfill\vrule height10pt width4pt depth2pt \par\hbox to \hsize{}}
\def\pref#1{(\ref{#1})}

\usepackage{amsmath}
\usepackage{amsfonts}
\usepackage{amssymb}
\usepackage{epstopdf}

\font\msytw=msbm9 scaled\magstep1 
scaled\magstep1

\let\a=\alpha \let\b=\beta  \let\g=\gamma  \let\d=\delta
\let\e=\varepsilon
\let\z=\zeta  \let\h=\eta   \let\th=\theta  \let\l=\lambda
\let\m=\mu    \let\n=\nu    \let\x=\xi         \let\r=\rho
\let\s=\sigma \let\t=\tau    
\let\ps=\Psi   \let\o=\omega
 \let\D=\Delta  \let\L=\Lambda

\def\EE{{\cal E}} \def\VV{{\cal V}}
 
\def\TT{{\cal T}} \def\BBB{{\cal B}}
\def\RR{{\cal R}}\def\LL{{\cal L}}

 \def\pp{{\bf p}}
 \def\xx{{\bf x}} \def\yy{{\bf y}}

\def\TTTT{\mathbf{T}}

\def\nn{\nonumber}

\def\RRR{\hbox{\msytw R}}

 \def\ZZZ{\hbox{\msytw Z}}
 
\def\TTT{\hbox{\msytw T}}

\def\\{\hfill\break}
\def\={:=}
\let\io=\infty
\let\0=\noindent

\def\tende#1{\,\vtop{\ialign{##\crcr\rightarrowfill\crcr\noalign{\kern-1pt
    \nointerlineskip} \hskip3.pt${\scriptstyle #1}$\hskip3.pt\crcr}}\,}
\def\otto{\,{\kern-1.truept\leftarrow\kern-5.truept\to\kern-1.truept}\,}

\def\wh{\widehat}
\def\to{\rightarrow}
\def\la{\left\langle}
\def\ra{\right\rangle}
\def\qed{\hfill\raise1pt\hbox{\vrule height5pt width5pt depth0pt}}

\def\be{\begin{equation}}
\def\ee{\end{equation}}
\def\bp{\begin{pmatrix}}
\def\ep{\end{pmatrix}}
\def\bea{\begin{eqnarray}}
\def\eea{\end{eqnarray}}
\def\nn{\nonumber}
\def\pref#1{(\ref{#1})}

\def\lb{\label}

\def\Tr{\mathrm{Tr}}

\newtheorem{theorem}{Theorem}[section]

\begin{document}

\title{ Vanishing of Drude weight in
interacting fermions on $\ZZZ^d$ with quasi-periodic disorder}

\author[1]{Vieri Mastropietro}
\affil[1]{University of Milano, Department of Mathematics ``F. Enriquez'', Via C. Saldini 50, 20133 Milano, Italy}

\maketitle
\begin{abstract} We consider a fermionic many body system in $\ZZZ^d$
with a short range interaction and quasi-periodic disorder.
In the strong disorder regime and assuming a Diophantine condition on the frequencies and on the chemical potential,
we prove at $T=0$ the exponential decay of the correlations 
and the vanishing of
the Drude weight, signaling Anderson localization in the ground state.
The proof combines Ward Identities, Renormalization Group
and KAM Lindstedt series methods.
\end{abstract}

\section{Introduction} 

The conductivity properties in fermionic systems, describing electrons in metals,
are strongly affected by the presence of disorder, which breaks the perfect periodicity of an ideal lattice and is unavoidable in real systems.
Disorder can be represented either by a random variable or by a quasi-periodic potential;
the first description is more suitable for impurities in solids while the second appears naturally in quasi-crystals or cold atoms experiments.
In absence of many body interaction 
disorder produces the phenomenon of Anderson localization \cite{A0}, consisting in an exponential decay of
all eigenstates and in an insulating behavior with vanishing conductivity. 
Such a phenomenon relies on the properties of the single particle Schroedinger equation
and it has been the subject of a deep mathematical investigation.
With random disorder Anderson localization was
established for strong 
disorder in any dimension \cite{loc0}, \cite{loc1} and 
in one dimension with any disorder. In the case of quasi-periodic disorder localization in one dimension is present only for large disorder
\cite{S}, \cite{FS}, while for weak disorder is absent; in higher dimensions localization was proved for strong disorder in $d=2$ 
\cite{b1}, \cite{b2} and for any $d$
in  \cite{1}.

The interplay between disorder and interaction has been deeply analyzed in the physical literature soon after \cite{A0}. The presence of many body interaction induces new processes which can indeed destroy localization.
At zero temperature $T=0$ with random disorder
qualitative scaling arguments 
gave evidence
of persistence of localization in $d=3$
\cite{A}, \cite{F} for short range weak interaction;  
in $d=1$ a second order Renormalization Group analysis was shown 
to produce a complex phase diagram \cite{Gi}.  The case of quasi-random disorder has been less studied, with the exception of 
\cite{M1a}, \cite{G1a} focusing on the extended weak disorder regime at $T=0$.
In more recent times 
the properties at $T>0$ were analyzed in 
\cite{Ba}, where perturbative arguments for the vanishing of conductivity
up to a certain critical $T$ in any dimension were given (many body localized phase).
Subsequently
numerical simulations found localization in certain systems in all the spectrum and vanishing of conductivity for any $T$, a phenomenon called
{\it many body localization}, see 
\cite{H4} for random and \cite{H5} for quasi-periodic disorder.
If all states are localized  
one expects, in a non-equilibrium setting, that interaction is unable to produce 
thermalization in an isolated quantum system, a phenomenon that in classical mechanics 
is due to closeness to an integrable system.  
Interacting quantum systems  with 
quasi-periodic 
disorder have been realized
in cold atoms experiments \cite{H19}, \cite{Z0},\cite{Z1}
; quasi-periodic disorder with many body interaction 
has been extensively numerically analyzed 
\cite{a}-\cite{h}. 

While the above works suggest that localization persists in presence of interaction,
results based on numerical or perturbative analysis cannot be conclusive.
In particular the presence of 
{\it small divisors} has the effect that 
physical informations are difficult to be extracted by lower order analysis but are typically
encoded in convergence or divergence of the whole series. This is a well known phenomenon in classical mechanics;
the Birkoff series for prime integrals in Hamiltonian systems
are generically diverging while Lindsdtet series for Kolomogorov-Arnold-Moser (KAM) tori
converge, even if both series are order by order finite and present similar small divisors. Therefore, even if 
perturbative analysis in \cite{Ba} or  \cite{Ro} 
get localization at finite temperature and in any dimension, one cannot exclude that the series are divergent and localization eventually disappear (this would say that thermalization
in experiments is eventually reached, even if at long times). 
A non-perturbative proof of many body localization
for all eigenstates has been indeed finally obtained in $d=1$ with random disorder in \cite{I} but
the result is based on a certain unproven assumption. A  complete proof 
have been obtained only with vanishing densities \cite{W}, \cite{W1}.
Arguments for breaking of many body localization in $d>1$
have been indeed presented in \cite{H6}.

In order to get rigorous results as benchmark for conjectures and approximations, 
a natural starting point is the zero temperature case in the thermodynamic limit. 
Our approach is to compute thermodynamical correlations; 
they not only provide
physical observables at equilibrium but give also information on the spectrum (so their computation is of interest even
for situation where equilibrium is not reached).
In particular at zero temperature they provide information of correlations over the ground state, 
while the vanishing of conductivity at any temperature is a signal of many body localization in all the spectrum. 
It has been proven 
in \cite{M1},\cite{M2},\cite{M3}
for one dimensional interacting fermions with strong quasi-periodic disorder
the $T=0$ exponential decay of 2-point correlations, indicating
 persistence of localization in the ground state.
Aim of this paper is twofold. 
The first is to investigate the $d>1$ case.
We consider a disorder of the form
$f(\vec \o \vec x)$ with $f$ periodic, as the one 
considered in \cite{b1} for the single particle Schroedinger equation
; more general forms of disorder are however possible, as
$f(\vec \o_1 \vec x,\vec \o_2 \vec x)$
considered in
\cite{b1}. The second aim is to compute the $T=0$ conductivity  
expressed by Kubo formula, whose properties
can be analyzed via a combination of information provided by Ward Identities with regularity properties
of the current correlations. The thermodynamical quantities are expressed by a series expansion
showing a peculiar combinations of properties appearing in classical and 
quantum physics; they show a small divisor problem, as in the Lindstedt series for KAM \cite{G1},
but loop graphs appear in the expansion, a signature of quantum physics totally absent in classical mechanics. In order to achieve convergence and exclude non perturbative effects one has from one side to show that divisors can be controlled by number theoretical conditions on frequencies, and 
from the other that the huge number of loop graphs is compensated  
by cancellations
from the fermionic anticommutative nature of the problem.

%

The paper is organized in the following way. In \S 2 the model is presented
and in \S 3 the main results, together with open problems, are presented. In \S 4 we discuss the implications of Ward Identities and regularity bounds. In \S 5 we introduce the Grassmann representation and in \S 6 we introduce the multiscale analysis. In \S 7 we prove the convergence 
of series expansion and in \S 8 we get the asymptotic decay of correlations.

\section{Interacting fermions with quasi-periodic disorder} 



We introduce the Fock space $\mathcal{F}_{L} = \bigoplus_{N\geq 0} \frak{h}_{L}^{\wedge N}$
where the $N$ particle Hilbert space $\frak{h}_{L}^{\wedge N}$ is the set of the totally antisymmetric square integrable functions
in 
$
\L_{L} :=\{ \vec x \in \mathbb{Z}^{d} \mid \vec x = n_{1} \vec e_{1} + n_{2} \vec e_{2}+...\;,\quad -L/2\leq n_{i} \leq L/2\;,\quad i=1,2,..,d\}$ 
where $\vec e_{i}$ are unit vectors.
The $a^{\pm}_{\vec x}$ are fermionic creation or annihilation operators
sending an element of $\frak{h}_{L}^{\wedge N}$ in $\frak{h}_{L}^{\wedge N+1}$ (creation) or  $\frak{h}_{L}^{\wedge N-1}$ (annihilation) and
$
\{ a^{+}_{\vec x}\,, a^{-}_{\vec y} \} = \delta_{\vec x,\vec y} $,
$\{ a^{+}_{\vec x}\, , a^{+}_{\vec y} \} = \{ a^{-}_{\vec x}\, , a^{-}_{\vec y} \} = 0$. The Hamiltonian is
\be
H=-{\e\over 2}
\sum_{\vec x}\sum_{i=1}^d (a^+_{\vec x+\vec e_i} a^-_{\vec x} +a^+_{\vec x} a^-_{\vec x+\vec e_i})
+u\sum_{\vec x} \phi_{\vec x}   a^+_{\vec x} a^-_{\vec x}+\l \sum_{\vec x}\sum_{i=1}^d
 a^+_{\vec x} a^-_{\vec x} a^+_{\vec x+\vec e_i} a^-_{\vec x+\vec e_i}\label{ll}
\ee
where $a^+_{\vec x}$ must be interpreted as zero for $\vec x\not\in \L_L$ and $\phi_{\vec x}=\bar\phi(\vec \o\vec x)$
with $\bar\phi(t):\TTT\to \RRR$ periodic of period $1$. In order to describe 
a quasi-periodic disorder we impose that $\vec \o$ is 
rationally independent and "badly" approximated by rationals (Diophantine condition).
The first term in \pref{ll} represents the kinetic energy of the fermions hopping on a lattice, the second represents the interaction with a quasi-periodic potential and the last term represents 
a 2 body interaction. 

There are several interesting limits; $\l=0$ is the non interacting limit;
$\l=u=0$ is the integrable limit;ùù
$\l=\e=0$ is the anti-integrable limit (the therminology was introduced in \cite{Au} ).
We consider the case in which $\l, \e$ are small with respect to $u$, and we set $u=1$
for definiteness; that is we consider a perturbation of the anti-integrable limit.

If 
$N=\sum_{\vec x} a^+_{\vec x} a^-_{\vec x}$ we define
\be
\langle \cdot \rangle_{\beta,L} = \frac{\Tr_{\mathcal{F}_{L}}\cdot e^{-\beta (H-\m N) }}{\mathcal{Z}_{\beta,L}}\;,\qquad \mathcal{Z}_{\beta,L} = \Tr_{\mathcal{F}_{L}}e^{-\beta
(H-\m N)}\label{11}\ee
where 
$\m$ is the chemical potential, which is fixed by the density in the Grand-Canonical ensamble, and  $\mathcal{Z}_{\beta,L}$ is the partition function. In the limit $\b\to\io$
they provide information on the ground states. We define
\be
\langle \cdot \rangle=\lim_{\b\to\io}  \lim_{L\to\io} \langle \cdot \rangle_{\beta,L}
\ee
The imaginary-time (or Euclidean) evolution of the fermionic operators is \be
a^{\pm}_{\xx} = e^{x_{0} (H-\m N)} a^{\pm}_{\vec x} e^{-x_{0}(H-\m N)}\ee 
with $\xx = (x_{0}, \vec x)\quad \text{with}\quad x_{0}\in [0, \beta)$,
The 2-point function is given by
%
\be
S_{\b,L}  (\xx,\yy)=\la T a^-_\xx a^+_\yy  \ra_{\b,L}
\ee
%
%
and $T$ is the time order product.  We also consider the truncated expectations  
$\la T A ;B\ra_{\b,L}=\la T A B\ra_{\b,L}-\la T A \ra_{\b,L}\la T B \ra_{\b,L}$.
The density and the current are given by
\be
\r_{\vec x}=a^+_{\vec x} a^-_{\vec x}\quad\quad
j^i_{\vec x}=
{\e\over 2 i}(a^+_{\vec x+\vec e_i} a^-_{\vec x}-a^+_{\vec x} a^-_{\vec x+\vec e_i})
\ee
The (Euclidean) conductivity density  in the zero temperature limit is defined by Kubo formula
\be
\s^i_{\vec y}=\lim_{ p_0\to 0}{1\over p_0} \lim_{ \b\to \io} \lim_{L\to \io}
[\sum_{\vec x\in \L_L}  \int_0^\b  dx_0  e^{i p_0 x_0}  
\la T j^i_{\vec x, x_0}; j^i_{\vec y , 0}\ra_{\b,L} 
+<\t^i_{\vec y}>_{\b,L} ]\label{ss}
\ee
where
\be
\t^i_{\vec y}= - {\e\over 2} (a^+_{\vec y+\vec e_i} a^-_{\vec y} +a^+_{\vec y} a^-_{\vec y+\vec e_i})
\ee
%

The conductivity can be equivalently expressed in terms of the 
Fourier transform which is, in the $\b\to\io, L\to\io$ limit , $i=1,,d$
\be
\hat H_{ii}(\pp,\vec y)=
\sum_{\vec x\in \L} \int_{\RRR} dx_0  e^{i \pp \xx}  <T j^i_{\vec x, x_00} ; j^i_{\vec y, 0}>\label{jj} \ee
and similarly we define $\hat H_{\m\n}(\pp,\vec y)$, with $\m=0,1,...d$ ($\m=0$
is the density and $\m=1,...,d$ the current component).
We can rewrite \pref{ss} as
\be
\s^i_{\vec y}=\lim_{ p_0\to 0}  \lim_{ \vec p\to 0}  
{1\over p_0} [\hat H_{ii}(\pp,\vec y) +<\t^i_{\vec y}>]
\ee
Finally the (zero temperature) Drude weight, see eg \cite{Po}, \cite{p} , is defined as
\be
D^i_{\vec y}=\lim_{p_0\to 0}  \lim_{ \vec p\to 0}   [\hat H_{ii}(\pp,\vec y) +<\t^i_{\vec y}>]
\ee
%
In a perfect metal at equilibrium the Drude weight is non-vanishing implying that the conductivity is infinite;
a vanishing Drude weight signals a non-metallic behavior. 

In the above 
definitions of conductivity the order in which the limits are taken is essential; already in the integrable limit $u=\l=0$ reversing the order of the limits one obtains
a zero result, while the Drude weight is indeed non vanishing as a consequence of the non-continuity
of the Fourier transform of the current correlation.

%

\section{Main result} 

In the anti-integrable limit $\l=\e=0$ the eigenvalues of the Hamiltonian are, $\vec x\in \L_L$
\be
H_0= \sum_{\vec x\in \L_L}  \bar\phi(\vec\o \vec x) n_{\vec x}   \quad\quad n_{\vec x} =0,1
\label{zzz}
\ee
and the single particle eigenfunctions have the form of $\d_{\vec x, \vec y}$. The 2-point function is given by
\be
g(\xx,\yy)=\d_{\vec x,\vec y} e^{(\phi_{\vec x}-\m)(x_0-y_0) } 
[\th(x_0-y_0){1\over 1+e^{\b (\phi_{\vec x}-\m)} }-
\th(y_0-x_0){e^{\b (\phi_{\vec x}-\m)} \over 1+e^{\b (\phi_{\vec x}-\m)} }]
\ee
which can be equivalently written as 
\be
g(\xx,\yy)=\d_{\vec x,\vec y}{1\over\b}\sum_{k_0={2\pi\over\b}(n_0+{1\over 2})}
e^{-ik_0(x_0-y_0)}  \hat g(\vec x,k_0) =\d_{\vec x,\vec y}\bar g(\vec x; x_0-y_0)
\ee
with 
\be
\hat g(\vec x,k_0) ={1
\over -i k_0+\phi_{\vec x} -\m}
\label{prop}
\ee
We define
\be
\m=\bar\phi(\a)
\ee
and the occupation number  on the ground state is 
$\th(\bar\phi(\vec \o\vec x)-\bar\phi(\a))$; the choice of $\m$ fixes the averaged density.
The conductivity is exactly vanishing as the is proportional to $\e$. The density correlation
is
\be
<\r_\xx;\r_\yy>=\d_{\vec x,\vec y}\bar g(\vec x; x_0-y_0)\bar g(\vec x; y_0-x_0)
\ee

We want to investigate what happens when we consider a non-vanishing hopping
$\e\not=0$ and interaction $\l\not=0$. 
As usual in small divisor problems, 
we need to impose a Diophantine condition on the frequencies $\vec \o$ of the quasi-periodic disorder
that is
\be
||(\vec \o \vec x)||_{\TTT}\ge C_0 |\vec x|^{-\t}\quad\quad \vec x\in \ZZZ^d/\vec 0\label{dio1}
\ee
$||.||$ being the norm on the one dimensional torus with period $1$; we require also a Diophantine condition on the chemical potential, that is
\be
||(\vec \o \vec x)\pm 2\a ||_{\TTT}
\ge C_0 |\vec x|^{-\t}\quad\quad \vec x\in \ZZZ^d/\vec 0\label{b}
\ee
%
The complementary of the set of numbers $\o,\a$ verifying
the diophantine conditions for some $C_0$ has measure $O(C_0)$, see eg \cite{La}.

In general the value of the chemical potential is modified by the interaction;
in order to fix the interacting chemical potential to the value $\bar\phi(\a)$ 
we choose the bare one to $\m=\bar\phi(\a)+\n$ with $\n$ chosen properly.

Our main result is the following

\begin{theorem}
 Assume that $\m=\bar\phi(\a)+\n$ and $\phi_x=\bar\phi(\vec \o \vec x)$ with $\bar \phi: \TTTT\to\RRR$, even, differentiable and such that $v_0=\partial\bar\phi(\a)\not=0$: 
in addition $\vec\o$ verifies \pref{dio1} and $\a$ verifies \pref{b}.
There exists $\e_0$ and a suitable choice of $\n=O(\e_0)$ such that,
for $|\l|\le |\e|\le\e_0$ in the zero temperature and infinite volume limit
\begin{enumerate} 
\item The 2-point correlation verifies, for any $N$
\be
|S(\xx,\yy)|\le |\log \D_{\vec x,\vec y}| C_N {e^{- {1\over 4}|\log | \e|| |\vec x-\vec y|}
\over 1+(\D_{\vec x,\vec y}
 |x_0-y_0| )^N}
\ee
with
\be
\D_{\vec x,\vec y}=(1+\min (|\vec x|, |\vec y|))^{-\t}
\ee
\item The density and current correlations verify 
\be
|H_{\m,\n}(\xx,\yy) |\le \D_{\vec x,\vec y}^{-4}   C_N {e^{- {1\over 4}|\log |\e|||\vec x-\vec y|}
\over 1+(\D_{\vec x,\vec y} |x_0-y_0| )^N}\label{assa1}
\ee
\item The Drude weight is vanishing \be D^i_{\vec x}=0\label{assa}
\ee
\end{enumerate}
\end{theorem}
The above result says that there is exponential decay in the coordinate difference
in the fermionic and current correlations, signaling localization in the ground state 
with quasi periodic potential of the form
$\bar \phi(\vec \o \vec x)$ in any dimension. Moreover the Drude weight at $T=0$
is vanishing, implying a non-metallic behavior.
This result is obtained assuming a Diophantine condition on 
the frequencies and on the chemical potential (or equivalently on the densities), see
\pref{b}. As the estimate of the  radius of convergence $\e_0$ is proportional to $C_0$
to some power, with fixed $\e, \l$ we get a large measure set of densities for which localization is present (but not on an interval). 

Information on the conductivity are obtained by combining the Ward Identities
following from the conservation of the current with regularity properties 
of the Fourier transform of the correlations, which are related to the decay
in the coordinate space. In the case of non-interacting fermions,
or for $1 d$ interacting fermions without disorder, the slow power law  decay of correlations implies a non vanishing Drude weight, see \cite{M4a}. In the present case,  the decay in space is exponentially fast but
the decay in the imaginary time 
has rate not uniform in $\vec x, \vec y$, 
due to the lack of translation invariance.  
As a consequence, we can deduce the vanishing of the Drude weight but not of the conductivity.

The analysis is based on an extension of the Lindstedt series approach to KAM
tori with exact Renormalization Group methods for fermions.
The correlations are expressed by a series expansion
showing a small divisor problem, as in the Lindstedt series for KAM,
in graphs with loops, which are a peculiarity
of quantum physics. Small divisors are controlled by the Diophantine conditions
and the huge number of loop graphs is compensated  
by cancellations due to anticommutativity.

While we have proved here the vanishing of the Drude weight, it would be interesting
to understand if also the conductivity is vanishing
or if a zero result is found only by a suitable averaging over the phase, as is  
done in numerical simulations \cite{g}. 

The effective interaction is irrelevant in the Renormalization Group sense, 
as consequence of Diophantine conditions and by cancellations due to anticommutativity.
The presence of spin \cite{M4} and an anisotropic hopping  \cite{M5}
produce extra marginal couplings. They can in principle destroy the convergence result of the
present paper, and it is interesting to observe that numerical 
\cite{Z} or cold atoms experiments \cite{Z1} have found evidence of delocalization is such cases.
Another important point would be to extend the analysis to a
more general kind of disorder like $f(\vec \o_1 \vec x, \vec\o_2 \vec x )$. 
The condition of strong disorder is non technical; in the case of weak quasiperiodic  disorder there is no localization; in particular, this is the case
of the interacting Aubry-Andre' model \cite{M6},
of the bidimensional Hofstadter model \cite{M7} or of three dimensional Weyl semimetals 
\cite{M8}.
Finally, we stress that a rigorous understanding of 
$T=0$ properties of interacting fermions with finite density and random disorder is still unknown.

The main open problem if of course to extend the above result on transport coefficients
to finite temperature to get information on localization beyond the ground state. 
While an extension of \cite{Po} allows to pass
from Euclidean to real time conductivity at $T=0$, this is expected to be a major 
difficulty for
$T>0$. Another difficulty is due to the fact that we do not get ground state 
localization in an interval of densities, but only in a large measure set.
The absence of thermalization in the classical case is considered related to KAM
theorem; it is interesting to note that the persistence of localization in a quantum
system, which is considered an obstruction to thermalization,
is also obtained via the generalization of KAM methods in a quantum context.

\section{Vanishing of Drude weight}

We show that the vanishing of Drude weight \pref{assa}
is consequence of the bound \pref{assa1}
combined with Ward Identities. Note first that the Fourier 
transform in the infinite volume limit is continuous as 
\bea
&&|\hat H_{\m,\n}(\pp,\vec y)|\le \sum_{\vec x} \int dx_0 
|H_{\m,\n}(\xx,\yy) |\le
\sum_{\vec x} \int dx_0 
\D_{\vec x,\vec y}^{-4}   C_N {e^{- {1\over 4}|\log |\e||\vec x-\vec y|}
\over 1+(\D_{\vec x,\vec y} |x_0| )^N}\le\\
&&C_1 \sum_{\vec x}   
 (|\vec x+\vec y|^{ 5\t}+|\vec y|^{5\t})
e^{- {1\over 4}|\log | \e||\vec x||}\le 
C_2 \sum_{\vec x}    e^{- {1\over 4}|\log |\e||\vec x||}
 (|\vec x|^{5\t} +2 |\vec y|^{5\t})\le 
C_3 |\vec y|^{5\t}/(|\log|\e||)^{d+5\t}\label{lips1}\nn
\eea
Ward identities can be deduced
from the continuity equation,
\be
\partial_0 \r_\xx=[H,\r_\xx]=-i \sum_i (j^i_\xx-j^i_{\xx-e_i})
\ee
we get, setting  $\partial_i  j_\xx\equiv  j_\xx-j_{\xx-{\bf e}_i}$ , $i=1,...,d$, ${\bf e}_i=(0, \vec e_i)$
\bea
&&\partial_0 <T \r_\xx;\r_\yy>=-i \sum_i \partial_i  <T j_\xx^i; \r_\yy>
+\d(x_0-y_0)<[\r_\xx,\r_\yy]> \nn\\
&&\partial_0 <T \r_\xx;j^j_\yy>=-i \sum_i \partial_i  <T j^i_\xx;j^j_\yy>+\d(x_0-y_0)
<[\r_\xx,j^j_\yy]>
\eea
Note that $[\r_{\vec x,x_0},\r_{\vec y,x_0}]=0$ while
\be
 [\r_{\vec x,x_0},j^j_{\vec y,x_0}]=-i \d_{\vec x, \vec y} \t^j_{\vec x}+i \d_{\vec x-\vec e_j, \vec y}\t^j_{\vec y}
\ee
so that, in the $L,\b\to\io$ limit
\bea
&&\partial_0 <T \r_\xx;\r_\yy>=-i \sum_i \partial_{i} <T j^i_\xx;\r_\yy>\label{al} \\
&&\partial_0 <T \r_\xx; j^j_\yy>=-i \sum_{i} \partial_{i}   <T j^i_\xx; j^j_\yy>-i\d(x_0-y_0)(
-\d_{\vec x, \vec y} <\t^j_{\vec y}>+\d_{\vec x-\vec e_j, \vec y}<\t^j_{\vec y}>)\nn
\eea
Taking the Fourier transform in $\xx$ we get, using translation invariance in time and setting $y_0=0$
\be
\sum_{\vec x} \int dx_0 e^{i \pp \xx}( \partial_0 <T \r_{\xx}; j^j_{\vec y}>+i\sum_{i} \partial_{i}   <T j^i_\xx ; j^j_{\vec y}>+i\d(x_0)(
-\d_{\vec x, \vec y} <\t^j_{\vec y}>+\d_{\vec x-\vec e_j, \vec y}<\t^j_{\vec y}>)=0
\ee
%
%
with $p_0\in \RRR$ and $\vec p\in [-\pi,\pi)^d$ so that
\be
-i p_0 \hat H_{0,j}(\pp,\vec y)+i \sum_i (1-e^{-i p_i})(\hat H_{i,j}
(\pp,\vec y)+
e^{-i \vec p \vec y}
 <\t^j_{y,0}>)=0\ee
Setting $j=1$ for definiteness, we set $\bar\vec p=(p_1,0,0)$ so that
\be
-i p_0 \hat H_{0,1}(\bar\pp,\vec y)+i(1-e^{-i p_1})(\hat H_{1,1}(\bar\pp,\vec y)+e^{-i p_1  y_1}
<\t^1_{y,y_0}>)
=0\ee
so that
%
%
%
\be
\lim_{p_1 \to 0}
(\hat H_{1,1}(0,p_1,\vec y)+e^{-i p_1  y_1}
<\t^1_{y,y_0}>)=0\ee
%
%
but $\lim_{p_1 \to 0}   (e^{-i p_1  y_1}-1)=0$.  In conclusion
\be
\lim_{p_1 \to 0}
(\hat H_{1,1}(0,p_1,\vec y)+
<\t^1_{y,y_0}>)=0
\ee
Due to \pref{lips1}
$\hat H_{1,1}(\pp,\vec y)$ is continuous in $\pp$ so that we can exchange the limits
\be
\lim_{p_0 \to 0}\lim_{\vec p \to 0}
(\hat H_{1,1}(\pp,\vec y)+<\t^1_{y,y_0}>)=D^1_{\vec x}=0
\ee
and this shows that the Drude weight is vanishing.
Note the crucial role played by continuity of the Fourier transform, following by the fast decay of the correlations; 
without quasi-periodic disorder the Fourier transform is not continuous due to its slow decay and the Drude weight is non vanishing.

\section{Perturbation theory and Grassmann representation} 

The starting point of the analysis consists in expanding around the anti-integrable limit \pref{zzz}; defining
\bea
&&H-\m N=H_0+V\\
&&H_0=
\sum_{\vec x} (\phi_{\vec x}-\bar\phi(\a))  a^+_{\vec x} a^-_{\vec x}\nn\\
&&V=
\e\sum_{\vec x,i} (a^+_{\vec x+\vec e_i} a^-_{\vec x} +a^+_{\vec x} a^-_{\vec x+\vec e_i})+\l \sum_{\vec x,i} a^+_{\vec x} a^-_{\vec x} a^+_{\vec x+\vec e_i} a^-_{\vec x+\vec e_i}+\n  \sum_{\vec x} a^+_{\vec x} a^-_{\vec x} \label{ll1}
\eea
and using the Trotter formula
%
%
one can write the partition function and the correlations as a power series expansion
in $\l,\e$.
%
%
%
\insertplot{550}{120}
{\ins{170pt}{90pt}{$\xx\pm{\bf e}_i$}
\ins{110pt}{90pt}{$\xx\pm{\bf e}_i$}
\ins{110pt}{20pt}{$\xx$}
\ins{180pt}{20pt}{$\xx$}
\ins{243pt}{30pt}{$\xx$}
\ins{283pt}{30pt}{$\xx\pm{\bf e}_i$}
\ins{343pt}{30pt}{$\xx$}
\ins{383pt}{30pt}{$\xx$}
\ins{370pt}{10pt}{$\n$}
\ins{270pt}{10pt}{$\e$}
\ins{150pt}{10pt}{$\l$}
}%
{verticiT333}
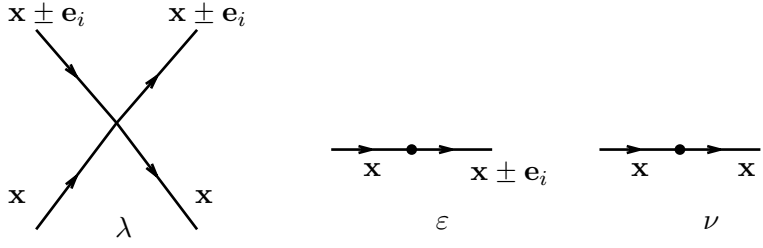
{\label{n9} Graphical representation of the three terms in $\VV(\psi)$ eq.\pref{VM}
}{0}
The correlations can be equivalently written in terms of Grassmann integrals. We can write
\be
e^{W(\h,J)}=\int P(d\psi)e^{-\VV(\psi)- \BBB(\psi,J,\h)}\label{GI}
\ee
with ${\bf e_i}=(0,\vec e_i)$
\be\VV(\psi)=\e \sum_i \int d\xx ( 
\psi^{+}_{\xx+{\bf e_i}}\psi^{-}_{\xx}
+\psi^{+}_{\xx-{\bf e_i}}\psi^{-}_{\xx})+
\l\int d\xx \sum_{i}
\psi^{+}_{\xx}\psi^{-}_{\xx} \psi^{+}_{\xx+{\bf e}_i}\psi^{-}_{\xx+{\bf e_i}}
+\n\int d\xx\label{VM}
 \psi^{+}_{\xx}\psi^{-}_{\xx}\ee
where $\int d\xx=\sum_{x\in\L_L}\int_{-{\b\over 2}}^{\b\over 2} dx_0$
and $\psi^\pm_\xx$ is vanishing outside $\L_L$; moreover
\be
\BBB(\psi,J,\h) =   
\int d\xx [\h^+_{\xx} \psi^-_{\xx} + \psi^+_{\xx}
\h^-_{\xx}+\sum_{\m=0}^d  J_\m(\xx) j_\m(\xx)]
\ee
with
\bea
&&j_0(\xx)=
\psi^+_{\xx}\psi^-_\xx\quad\quad j_i(\xx)=\e(
\psi^+_{\xx+{\bf e}_i}\psi^-_\xx-\psi^+_{\xx}\psi^-_{\xx+{\bf e_i}})
\eea
The 2-point and the current correlations are given by
\be
S_{2}^{L,\b}(\xx,\yy)
={\partial^2\over\partial\h^+_{\xx}\partial\h^{-}_{\yy}}
W(\h,J)|_{0,0}\quad\quad H_{\m,\n}(\xx,\yy)={\partial^2\over\partial J_{\m,\xx}\partial J_{\n,\yy}}
W(\h,J)|_{0,0}
\label{asso}
\ee
By expanding in $\l,\e,\n$ one can write the correlations as a series expansion, which
can be expressed in terms of Feynman graphs obtained contracting the half lines of vertices, see Fig. 1, and associating to each line the propagator $g(\xx,\yy)$.
There is a basic difference between the perturbative expansion in the non interacting case $\l=0$ and the interacting case $\l\not=0$. In the first case there are only chain graphs,
while in the second there are also loops, producing further combinatorial problems.
One can verify that the perturbative expansions obtained by Trotter formula for
\pref{11} and by the Grassmann generating functions are the same (this is true up to the 
so called "tadpoles" which can be easily taken into account, see \S  1 D in \cite{M2}). 
The identity between \pref{11}
and \pref{GI}
is true in a rigorous sense provided that 
the Grassmann integral representation is analytic in a disk
uniformly in $L,\b$, as proven in the following sections.
Indeed
at finite $L,\b$ the partition function in \pref{11}
 is entire and it coincides order by order
 with the Grassmann representation, which is analytic in a disk independent on the volume, 
so they coincide. As the denominator of the correlations
is non vanishing in this finite disk and the numerator is entire at finite $\b,L$, also the correlations
\pref{11} is analytic and
 coincide with the Grassmann representation, and the identity holds also in the limit.

\section{Multiscale decomposition and renormalization} 

The difficulty in controlling the perturbative expansion is due to a "small divisor problem"
related to the size of the propagator; the denominator of $\hat g(\vec x, k_0)$
can be arbitrarily small if $\vec \o\vec x$ is 
close to $\pm \a$, a fact which can produce in principle $O(n!)$-terms which could destroy convergence.
The starting point of the analysis is to 
separate the propagator in two terms,
one containing the quasi-singularity and a regular part; we write
\be
g(\xx,\yy)=g^{(1)}(\xx,\yy)+\sum_{\r=\pm} g_\r^{(\le 0)}(\xx,\yy)
\ee
where
\bea 
&&g^{(1)}(\xx,\yy)={\d_{\vec x,\vec y}\over \b}\sum_{k_0}\chi^{(1)}(\vec\o\vec x, k_0)
{e^{-i k_0(x_0-y_0)}\over -i k_0+\bar\phi (\vec\o\vec x) -\bar\phi (\a)
}=\d_{\vec x,\vec y}g^{(1)}(\vec x, x_0-y_0)    \nn\\
&&g^{(\le 0)}_\r(\xx,\yy)={\d_{\vec x,\vec y}\over \b}\sum_{k_0}\chi^{(0)}_\r(\vec\o\vec x,k_0)
{e^{-i k_0(x_0-y_0)}\over -i k_0+\bar\phi (\vec\o\vec x) -\bar\phi (\a)
}=\d_{\vec x,\vec y}g^{(\le 0)}_\r(\vec x,x_0-y_0)  \label{assd}
\eea
with $\chi^{(0)}_\r(\vec\o\vec x,k_0)=
\tilde\th_\r(\vec\o\vec x)
\bar\chi_0(\sqrt {k_0^2+(\bar\phi(\vec\o \vec x)- \bar\phi(\a))^2})$
with $\tilde\th_\r$ is the periodic theta function ($\tilde
\th_\pm=1$ if $\vec\o\vec x$ mod. $1$
is positive/negative and zero otherwise)    
and $\bar\chi_0$ such that $C^\io(\RRR^+)\to \RRR$  
such that $\bar\chi_0(t)=1$ 
with $t\le 1$ and  $\bar\chi_0(t)=0$ for $t\ge \g>1$;  moreover 
$\chi^{(1)}+\sum_{\r=\pm} \chi_\r=1$. The "infrared" propagator $g^{(\le 0)}(\xx,\yy)$ has denominator arbitrarily small. We can further decompose the infrared propagator as
sum of propagators with smaller and smaller denominators
\be
g^{(\le 0)}_\r(\vec x, x_0-y_0)=  \sum_{h=-\io}^0  g^{(h)}_\r(\vec x, x_0-y_0)
\ee
with $g^{(h)}_\r$ similar $g^{(\le 0)}_\r$ witrh $f^h$ replacing $\bar\chi_0$ with
\be
f^h=\bar\chi_0(\g^h\sqrt {k_0^2+(\bar\phi(\vec\o \vec x)- \bar\phi(\a))^2})-\bar\chi_0(\g^{h-1}\sqrt {k_0^2+(\bar\phi(\vec\o \vec x)- \bar\phi(\a))^2})
\ee
For any integer $N$ one has
\be
|g^{(h)}_\r(\vec x, x_0-y_0)|\le {C_N\over 1+(\g^h|x_0-y_0|)^N }
\ee
if $C_N$ is a suitable constant.

The integration of \pref{GI}
is done iteratively by using two crucial properties of Grassmann integrations.
%
%
If $P(d\psi^{(1)})$ and $P(d\psi^{(\le 0)})$ are gaussian Grassmann integrations
with propagators $g^{(1)}$ and $g^{(\le 0)}$, we can write $P(d\psi)
=P(d\psi^{(1)})P(d\psi^{(\le 0)})$ so that
\bea
&& e^{W(\h,J)}=
\int P(d\psi^{(1)})P(d\psi^{(\le 0)})
e^{-\VV(\psi^{(1)}+\sum_{\r=\pm} \psi^{(\le 0)}_\r )- \BBB(\psi^{(1)}+\sum_{\r=\pm} 
\psi^{(\le 0)}_\r,\h,J)}=\nn\\
&&\int P(d\psi^{(\le 0)})
e^{-\VV^{(0)} (\psi^{(\le 0)}_\r,\h,J)}
\label{GI1}
\eea
with 
\be
\VV^{(0)} (\psi^{(\le 0)}_\r,\h,J)=\sum_{n=0}^\io {1\over n!} \EE^T_1(\VV+\BBB;n)
\ee
and  $\EE^T_1$ are fermionic truncated expectations with propagator $g^{(1)}$.
By integrating $\psi^{(0)}, \psi^{(-1)},..,\psi^{(h+1)}$ one obtains a sequence of 
effective potentials $\VV^{(h)}$, $h=0,-1,-2,..$. The
way in which we define the integration is dictated by the scaling dimension which is, as we will see below, $D=1$; that is all terms are relevant in the Renormalization Group sense.
\vskip.3cm
{\bf Remark} Note that after the integration of $\psi^1$ one gets a theory defined in terms of two fields $\psi_+,\psi_-$. This is due to the fact that $\bar\phi(t)=\bar\phi(\a)$ in correspondence of
two points $\pm \a$. If we consider more general forms of quasi periodic disorder, like $\bar\phi(t_1,t_2)$ as the one in \cite{b2}
, then $\bar\phi(t_1,t_2)-\m=0$ in a set corresponding to a surface.
In this case one gets a description in terms of a field $\psi_\r$, with $\r$ a
parameter parametrizing this curve, a situation 
somewhat analogue to what happens in interacting fermions with extended Fermi surface.

\vskip.3cm
The multiscale integration is described iteratively in the following way.
Assume that we have already integrated the fields 
$\psi^{(0)}, \psi^{(-1)},..,\psi^{(h+1)}$ obtaining (we set $\h=0$ for the moment)
\be
 e^{W(0,J)}=\int P(d\psi^{(\le h)})e^{-\VV^{(h)}(\psi^{(\le h)},J)}\label{gr2} \ee
where $P(d\psi^{(\le h)}$ has propagator 
\be
g^{(\le h)}_\r(\xx,\yy)={\d_{\vec x,\vec y}\over \b}\sum_{k_0}\chi^{(h)}_\r(k_0,\vec\o\vec x)
{e^{-i k_0(x_0-y_0)}\over -i k_0+\bar\phi (\vec\o\vec x) -\bar\phi (\a)
}=\d_{\vec x,\vec y}g^{(\le 0)}_\r(\vec x,x_0-y_0)  
\ee
and 
\be \VV^{(h)}(\psi^{(\le h)},J)=\sum_{l\ge 0,m\ge 0}\sum_{\underline\e,\underline\r}
\int d\xx_{1}...d\xx_{l} d\yy_{1}...d\yy_{m}
H^h_{l,m}(\underline \xx,\underline \yy)
\prod_{i=1}^{l} \psi^{\e_i (\le h)}_{\r_i,\xx_i}
\prod_{i=l}^m  J_{\yy_i}   
\ee
%
%
%

If there is a subset of $\psi^{\e_i}_{\r_i,\xx_i}$ with the same $\e,\r$ and $\vec x_i$,
by the anticommuting properties of Grassmann variables we can write, if $l>1$
\be
\prod_{i=1}^{l} 
\psi^{\e}_{\vec x,x_{0,i}}=\psi^{\e}_{\vec x,x_{0,1}}\prod_{i=2}^l D^{\e}_{\vec x, x_{0,i},x_{0,1} }\quad\quad \quad D^{\e}_{\vec x, x_{0,i},x_{0,1} }=\psi^{\e}_{\vec x,x_{0,i}}-\psi^{\e}_{\vec x,x_{0,1}}
\ee
We can therefore rewrite that effective potential in the following way
\be \VV^{(h)}(\psi^{(\le h)},J)=\sum_{l\ge 0,m\ge 0}\sum_{\underline\e,\underline\r}
\int d\xx_{1}...d\xx_{l} d\yy_{1}...d\yy_{m}
H^h_{l,m}(\underline \xx,\underline \yy)
\prod_{i=1}^{l}  d^{\s_i}\psi^{\e_i}_{\r_i,\xx_i}
\prod_{i=l}^m  J_{\yy_i}   
\ee
with $\s=0,1$ and 
$d^{0}\psi=\psi$ and $d^{1}\psi=D$.

We define {\it resonant} the terms with fields with the same coordinate $\vec x$, that is 
$\xx_i=(x_{0,i},\vec x )$.
Note that all the resonant terms with $l\ge 4$ are such that 
there are at least two $D$ fields; 
the fields have the same $\r$ index as have the same 
$\vec \o \vec x$.

We define a {\it renormalization operation} $\RR$ in the following way
%
%
%
\begin{enumerate}
\item If $l=2$, $m=0$ 
%
%
\be \RR \sum_{\vec x}\int dx_{0,1} dx_{0,2}
H_{2,0}^{(h)} 
\psi^{+(\le h)}_{\vec x,x_{0,1}, \r}\psi^{-(\le h)}_{\vec x,x_{0,2}, \r}
=\sum_{\vec x}\int dx_{0,1} dx_{0,2}H_{2,0}^{(h)}  
\psi^{+(\le h)}_{\vec x,x_{0,1}, \r}T^{-(\le h)}_{\vec x,x_{0,1},x_{0,2} \r}
\ee
with
\be
T^{-(\le h)}_{\vec x,x_{0,1},x_{0,2} \r}=
\psi^{-(\le h)}_{\vec x,x_{0,2}, \r}
- \psi^{-(\le h)}_{\vec x,x_{0,1}, \r}-(x_{0,1}-x_{0,2}) 
\partial\psi^{-(\le h)}_{\vec x,x_{0,1}, \r}
\label{T}
\ee
\item $\RR=0$ otherwise
%
%
%
%
%
%
\end{enumerate}
We define $\RR=1-\LL$ and  
by definition $\LL\VV^{(h)}$ is given by the following expression
\be \LL \VV^{(h)}= \g^h F^{(h)}_\n+  F^{(h)}_{\z}
+F^{(h)}_{\a}
\ee
where, if $H_{2,0}^{(h)}(\vec x, x_0-y_0)\equiv 
\bar H_{2,0}^{(h)}(\vec\o\vec x, x_0-y_0)$ one has 
 \be\n_h=\int dx_0 \bar H_{2,0}^{(h)}(\r \a, x_0)\quad
\x_h(\vec x)=\int dx_0 {\bar H_{2,0}^{(h)}(\vec \o \vec x, x_0)-
\bar H_{2,0}^{(h)}(\r \a, x_0)\over \vec \o \vec x-\r \a}
\ee
and 
$\a_h(\vec x)=\int dx_0 x_0 \bar H_{2,0}^{(h)}(\vec \o \vec x, x_0)$; moreover
\bea &&F^{(h)}_\n= \sum_{\r}\sum_{\vec x}\int dx_0 \n_h \psi^{+(\le
h)}_{\xx,\r}
\psi^{-(\le h)}_{\xx,\r}\quad\quad F^{(h)}_\z= \sum_{\r}\sum_{\vec x} \int dx_0
 ((\vec \o \vec x)-\r \a) \z_{h,\r}(\vec x)\psi^{+(\le
h)}_{\xx,\r}
\psi^{-(\le h)}_{\xx,\r}\label{zak}\nn\\
&&F^{(h)}_\a= \sum_{\r}\sum_{\vec x} \int dx_0  \a_{h,\r}(\vec x)\psi^{+(\le
h)}_{\xx,\r}\partial_0
\psi^{-(\le h)}_{\xx,\r}\quad\quad
\eea
%
%
%
The {\it running coupling constants } $\vec v_h=(\n_h,\a_h,\x_h)$ are {\it independent} from $\r$, 
as
\pref{GI} is invariant under parity $\vec x\to -\vec x$.
Note also that $(\hat g^{(k)})^*(\vec x,k_0)=\hat g^{(k)}(\vec x,-k_0)$ so that $(\hat H^{(h)}_{2,\r}(\vec x,k_0))^*=
\hat H^{(h)}_{2,\r}(\vec x,-k_0)$, and this implies that $\n_h$ is real. 

\vskip.3cm
{\bf Remark} The $\RR$ operation is defined in order to act non trivially on the resonant terms with two fields and no $J$ fields; they are the only resonant terms with no $D$ fields.
This fact would be not true of there is the spin 
or an extra degree of freedom, as in the case of lattice Weyl semimetals \cite{M8}.
In that case the local part of the effective potential would contain also effective interactions.
\vskip.3cm
With the above definitions we can write \pref{gr2}
\be
 e^{W(0,J)}=\int P(d\psi^{(\le h-1)})\int P(d\psi^{(h)})
e^{-\LL\VV^{(h)}(\psi^{(\le h)},J)-\RR\VV^{(h)}(\psi^{(\le h)},J)
}=\int P(d\psi^{(\le h-1)})\
e^{-\LL\VV^{(h)}(\psi^{(\le h-1)},J)}
\label{gr3} \ee
and the procedure can be iterated.

\section{Convergence of series expansion}

The effective potential can be written as a sum over Gallavotti trees
 $\t$, see Fig.2 
\be
\VV^{(h)}(\psi^{(\le h)},J) =
\sum_{n=1}^\io\sum_{\t\in\TT_{h,n}}
V^{(h)}(\t,\psi^{(\le h)})
\ee
where $\t$ are trees  constructed adding labels to the unlabeled trees,
obtained by joining a point, the {\it root}, with an ordered set of
$n\ge 1$ points, the {\it endpoints}, so that the root is not a branching point. 

\insertplot{300}{200}
{\ins{60pt}{90pt}{$v_0$}\ins{120pt}{100pt}{$v$}
\ins{100pt}{90pt}{$v'$}
\ins{120pt}{-5pt}{$h_v$}
\ins{235pt}{-5pt}{$1$}
\ins{255pt}{-5pt}{$2$}
}
{treelut2}{\label{n11} A labeled tree 
}{0}


The set of labeled trees $\TT_{h,n}$
is defined associating a label $h\le 0$ with the root and 
introducing
a family of vertical lines, labeled by an integer taking values
in $[h,2]$ intersecting all the non-trivial vertices, the endpoints and other points called trivial vertices.To a 
vertex $v$ is associated
$h_v$ and, if $v_1$ and $v_2$ are two vertices and $v_1<v_2$, then
$h_{v_1}<h_{v_2}$.
Moreover, there is only one vertex immediately following
the root, which will be denoted $v_0$ and can not be an endpoint;
its scale is $h+1$. To the end-points are associated $\VV+\BB$ , and in such a case
the scale is $2$;  or 
$\LL\VV^{h_v-1}(\psi^{(\le h_v-1)},J)$ and in this case the scale is $h_v \le 1 $ 
and there is the constraint 
that $h_v=h_{\bar v}+1$, if $\bar v$ is the first non trivial vertex immediately preceding $v$. The tree structure induces a jerarchy of end-points which can be represented by clusters, see Fig.3.

\insertplot{340}{160}
{\ins{125pt}{140pt}{$1$}
\ins{125pt}{100pt}{$2$}
\ins{125pt}{80pt}{$3$}
\ins{125pt}{60pt}{$4$}
\ins{125pt}{20pt}{$5$}
\ins{145pt}{80pt}{$\Longleftrightarrow$}
\ins{193pt}{75pt}{$1$}
\ins{237pt}{75pt}{$2$}
\ins{253pt}{75pt}{$3$}
\ins{261pt}{75pt}{$4$}
\ins{291pt}{75pt}{$5$}}
{fig10}
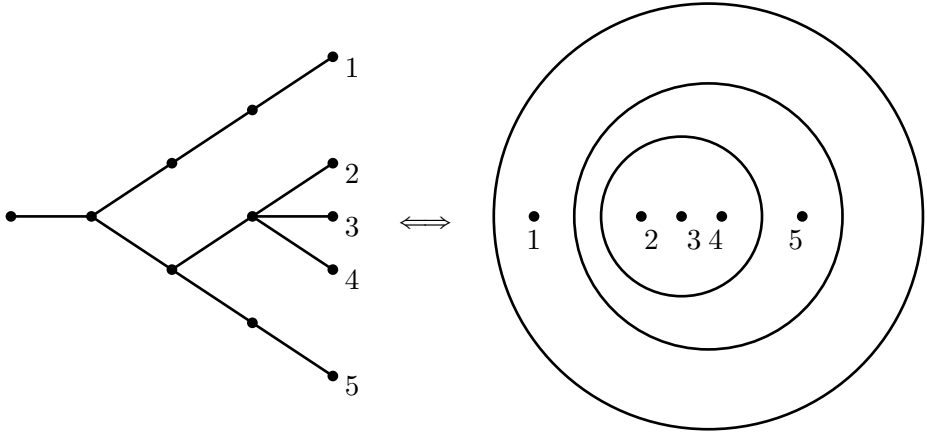
{\label{n10} A tree of order 5 and the corresponding
clusters.}{0}

If $v_0$ is the first vertex of $\t$ and $\t_1,..,\t_s$ ($s=s_{v_0}$)
are the subtrees of $\t$ with root $v_0$,\\
$V^{(h)}(\t,\psi^{(\le h)})$ is defined inductively by the relation
\be
V^{(h)}(\t,\psi)=
{(-1)^{s+1}\over s!} \EE^T_{h+1}[\bar
V^{(h+1)}(\t_1,\psi^{(\le h+1)});..; \bar
V^{(h+1)}(\t_{s},\psi^{(\le h+1)})]\label{3.33}
\ee
where $\bar V^{(h+1)}(\t_i,\psi^{(\le h+1)})$
it is equal to $\RR\VV^{(h+1)}(\t_i,\psi^{(\le h+1)})$ if
the subtree $\t_i$ is non trivial;if $\t_i$ is trivial, it is equal to $\LL\VV^{(h+1)}$.
By iterating \pref{3.33} we get a jerarchy of truncated expectations, with
a certain subset of fields contracted in each  expectations. We can therefore write
$V^{(h)}(\t,\psi^{(\le h)})$ as sum over sets defined in 
the following way. We call $I_v$ the set of $\psi$ associated to the end-points following $v$
and $P_v$ is a subset of $I_v$ denoting  the external $\psi$.
We
denote by $Q_{v_i}$ the intersection of $P_v$ and $P_{v_i}$; they are such that
$P_v=\cup_i Q_{v_i}$ and the union ${\cal
I}_v$ of the subsets $P_{v_i}\setminus Q_{v_i}$ is, by definition,
the set of the {\it internal fields} of $v$, and is non empty if
$S_v>1$.  The effective potential can be therefore written as
\be \VV^{(h)}(\t,\psi^{(\le h)})=\sum_{{\bf P}\in{\cal
P}_\t} 
\VV^{(h)}(\t,{\bf P})\quad \bar\VV^{(h)}(\t,{\bf P} )=\int d\xx_{v_0} \widetilde\psi^{(\le
h)}(P_{v_0}) K_{\t,{\bf P}}^{(h+1)}(\xx_{v_0})\;,\lb{2.43a}\ee
where $\widetilde\psi^{(\le
h)}(P)=\prod_{f\in P}\psi_{\xx(f)}$. 
If we expand the truncated expectations by the Wick rule we get a sum of Feynman graphs with an associated
cluster structure; an example is in Fig.4.
\insertplot{370}{150}
{
}
{figjsp467aa1}
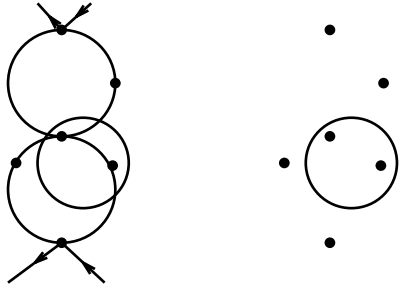
{\label{h2} An example of graph with $\l$ and $\e$ vertices and the associated 
cluster structure; the propagator in the cluster, represented as a circle,
has scale $h$ smaller than the scales of the propagators  external to the cluster.
} {0}

The truncated expectations can be written by the Brydges-Battle-Federbush formula
\be \EE^T_{h_v}(\tilde\psi^{(h_v)}(P_{1}/Q_1 ),\cdots,
\tilde\psi^{(h_v)}(P_{s}/Q_s)
))=\sum_{T_v}\prod_{l\in T_v}
\big[\d_{\vec x_l,\vec y_l}\bar g^{(h_v)}(\vec x_l,x_{0,l}-y_{0,l})\big]\, \int dP_{T}({\bf t})\; {\rm
det}\, G^{h_v,T}({\bf t})\;,\label{2.46aa}\ee
where $T_v$ is a set of lines forming an {\it anchored tree graph} between the
clusters of points $\xx^{(i)}\cup\yy^{(i)}$,
that is $T_v$ is a set of lines,
which becomes a tree graph if one identifies all the points in the same
cluster. Moreover ${\bf t}=\{t_{ii'}\in [0,1], 1\le i,i' \le s\}$,
$dP_{T_v}({\bf t})$ is a probability measure with support on a set of ${\bf t}$
such that $t_{ii'}={\bf u}_i\cdot{\bf u}_{i'}$ for some family of vectors
${\bf u}_i\in \RRR^{s}$ of unit norm.
\be G^{h,T}_{ij,i'j'}=t_{ii'}
\d_{\vec x_{ij},\vec y_{i'j'}} \bar g^{(h)}(\vec x_{ij}, x_{0,ij}-y_{0,i'j'})\;,
\label{2.48}\ee
%
%
%
%
%
We define $\bar T_v=\bigcup_{w\ge v}T_w$ starting
from $T_v$ and attaching to it the trees $T_{v_1},..,T_{v_{S_v}}$ associated to the vertices $v_1,..,v_{S_v}$
following $v$ in $\t$, and repeating
this operation until the end-points of $\t$ are reached.

\insertplot{200}{195}
{\ins{100pt}{110pt}{$w_1$}
\ins{75pt}{100pt}{$w_a$}
\ins{45pt}{80pt}{$w_b$}
\ins{20pt}{80pt}{$w_c$}
\ins{40pt}{40pt}{$w_2$}
}
{fig60}
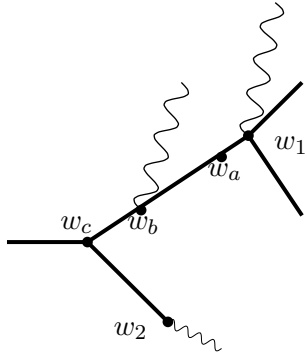
{\label{h2} A tree $\bar T_v$ with attached wiggly lines representing the 
 external lines $P_v$; the lines
represent propagators with scale $\ge h_v$ connecting $w_1,w_a,w_b,w_c,w_2$, representing
the end-points following $v$ in $\t$.  
} {0}

The tree $\bar T_v$ connects the end-points $w$ of the tree $\t$. To each end-point $w$ we associate
a factor $\vec\d_{w}^{i_w}$, and a) $\vec\d^i_w=0$ if $w$ corresponds to a $\n_h,\a_h,\z_h$ end-point;
b) $\vec\d_w^i$ one among $\pm \vec e_i$, $i=1,2,3$ if it corresponds to an $\e$ end-point;
c) $\d^i_w$ one among $0,\pm \vec e_i$, $i=1,2,3$
if it corresponds to a $\l$ end-point. If $\vec x_{w_1}$ and $\vec x_{w_2}$ are
coordinates of the external fields $\tilde\psi(P_v)$ we have, see Fig.5 
\be \vec x_{w_1}-\vec x_{w_2}=  \sum_{w\in c_{w_1,w_2}}
\vec\d_{w}^{i_{w}}\label{fa}
\ee
where $c_{w_1,w_2}$ is the set of endpoints in the path in $\bar T$ connecting $w_1$ and $w_2$.
The above relation implies, in particular, that the coordinates of the external fields $\tilde\psi(P_{v})$ are determined once that the choice of a single one of them and of $\t, \bar T_{v}$ and ${\bf P}$ is done. 
We can therefore write the effective potential as sum over trees $T$, setting the 
Kronecker
deltas in the propagators in $l\in  T$
equal to $1$
\be \VV^{(h)}(\t,\psi^{(\le h)})=\sum_{{\bf P}\in{\cal
P}_\t} \sum_{T}
\VV^{(h)}(\t,{\bf P},T)\quad \bar\VV^{(h)}(\t,{\bf P}, T )=\sum_{\vec x} \int dx_{0, v_0} \widetilde\psi^{(\le
h)}(P_{v_0}) K_{\t,{\bf P},T}^{(h+1)}(\xx_{v_0})\;,\lb{2.43a}\ee
where in $K_{\t,{\bf P},T}^{(h+1)}$ the propagators in $T$ are $g^{(h)}(\vec x, x_0-y_0)$
and the determinants are product of determinats involving propagators with the same $\vec x$.
We can bound the propagators in $T$ by
\be
\int dx_0 |g^{(h)}(\vec x, x_0-y_0)|\le C \g^{-h}
\ee
Moreover the determinants in the BFF formula can be bounded by the Gram-Hadamard inequality .
We introduce an Hilbert
space $\HH=\RRR^s\otimes L^2(\RRR^1)$ so that
\be
\tilde G^{h,T}_{ij,i'j'}=
\Big({\bf u}_{i}\otimes
A(x_{0, ij}-,x_{ij})\;,\ 
{\bf u}_{i'}
\otimes B(y_{0,i'j'}-,x_{ij})\Big)
\label{as}
\;,\ee
where 
${\bf u}\in \RRR^{s}$ are unit vectors $(u_i,u_{i})=t_{ii'}$, and $A,B$
\be (A,B)=\int dz_0 A(\vec x,x_0-z_0)B^*(\vec x,z_0-y_0)\ee
given by
\be A(\vec x,x_0-z_0)={1\over\b}\sum_{k_0} e^{-i k_0
(x_0-z_0)}\sqrt{f_h}\quad\quad 
B(\vec x,y_0-z_0)={1\over \b}\sum_{k_0} {e^{-ik_0( y_0-z_0)}\sqrt{f_h}
\over -i k_0+\bar\phi(\vec\o\vec x)-\bar\phi(\a)}\nn
\label{2.48b}\ee
%
%
%
%
Moreover
$||A_h||^2=\int dz_0 |A_h(x',z_0)|^2\le C\g^{h}$ and $||B_h||^2\le C \g^{-h}$ so that
By Gram-Hadamard inequality we get:
\be |{\rm det} \tilde G^{h_v,T_v}({\bf t}_v)| \le
C^{\sum_{i=1}^{S_v}|P_{v_i}|-|P_v|-2(S_v-1)}\;.\lb{2.54a}\ee
%
One get therefore the bound, for $|\l|, |\vec v_h|\le \e_0$, 
\be
|K_{\t,{\bf P},T}^{(h+1)}(\xx_{v_0})|\le  C^n\e_0^n \prod_v {1\over S_v!}\g^{-h_v(S_v-1)}\label{sss}
\ee
which is not suitable for summing over $\t$ and $P$.
In order to improve the above bound we need to implement in the bounds 
some constraints which have been neglected in the derivation of \pref{sss}, and to take into account the effect
of the presence of the $D$ fields. 

We define $V_\chi$ the set of non trivial vertices or the trivial ones with non zero internal lines;
we define $v'$ the first vertex in $V_\chi$ following $v$. 
We say that $v$ is a non-resonant vertex if in $\tilde\psi (P_v)$ there are at least two different coordinates, and 
a resonant vertex when all coordinates are equal. We define $S_v=S^L_v+S^H_v$ where $S^L_v$ is the number of non resonant subtrees (including trivial ones)  and 
 $S^H_v$  the number of resonant ones (inluding trivial ones). We also call $H$ the set of $v\in V_\chi$
which are resonant and $L$ the $v\in V_\chi$ which are non resonant.
Consider a non resonant vertex $v$ so that
there are at least two fields in $P_v$ with different spatial coordinates $\vec x$, say 
$\vec x_{w_1} \not =\vec x_{w_2}$. 
The fields $\tilde\psi^{(\le h_v)}(P_v)$ have scale $\le \g^{h_{v'}}$, $v'\in V_\chi$ the first vertex belonging to $V_\chi$ after $v$ 
so that
\be
||(\vec\o \vec x_{w_1})-\r_1 \a||_{\TTT}
\le c v_0^{-1} \g^{h_{v'}-1} \quad\quad
||(\vec\o \vec x_{w_2})-\r_2 \a||_{\TTT}\le c v_0^{-1} \g^{h_{v'}-1}
\ee
so that
\be
2 c v_0^{-1}\g^{h_{ v'}}\ge ||(\vec\o \vec x_{w_1})-\r_1 \a||_{\TTT}
+ ||(\vec\o\vec x_{w_2})-\r_2 \a||_{\TTT}\ge 
||\vec \o(\vec x_{w_1}-\vec x_{w_2})-(\r_1-\r_2)\a ||_{\TTT}
\ee
and by \pref{fa}
\be
2 c v_0^{-1}\g^{h_{ v'}}\ge ||\vec \o(\sum_{w\in c_{w_1,w_2}}
\vec \d_{w}^{i_{w}}) +(\r_1-\r_2)\a ||_{\TTT}
\ge {C_0\over |\sum_{w\in c_{w_1,w_2}}
\vec \d_{w}^{i_{w}}|^{\t}}
\ee
where the Diophantine conditions have been used. Therefore
\be
\sum_{w\in c_{w_1,w_2}}
|\vec\d_{w}^{i_{w}}|\ge |\sum_{w\in c_{w_1,w_2}}
\vec\d_{w}^{i_{w}}|\ge C \g^{-h_{ v'}/\t}
\ee
and, if $N_v$  is the number of end-points following $v$ in $\t$
\be
\sum_{w\in c_{w_1,w_2}}
|\vec\d_{w}^{i_{w}}|\le N_v
\ee
as $|\vec\d_{w}^{i_{w}}|=0,1$ so  that
\be
N_v\ge C \g^{-h_{ v'}/\t}\label{0}
\ee
Note that to each endpoint is associated a small factor $\e_0$ and the fact that $N_v$ is large by \pref{0}
produces a gain for the $v$ with the fields with different $\vec x$. 
Of course there can be several $\bar T_v$ with different $v$
passing through the same end-points. Therefore, 
given a constant $c<1$, we can multiply the contribution to each tree $\t$ with $n$-endpoints by 
$c^{-n} c^n$ (the factor $c^{-n}$ is of course armless); we can then write
\be
c=\prod_{h=-\io}^0 c^{2^{h-1}}\ee
and  associate to each $v$ a factor $c^{N_v 2^{h-1}}$. If 
there are
two fields in $P_v$ (that is external to the cluster $v$) with different $\vec x$
 we get in the bounds, by assuming $\g^{1\over\t}/2\equiv \g^\h>1$ than, for any $N$
\be
c^{A \g^{-h\over \t}2^{h}}=e^{-|\log c|
A \g^{-\h h}}\le \g^{N \h h} {N\over [|\log |c|| A ]^N e^{N}}\label{sap} 
\ee
as  $e^{-\a x} x^N\le [{N\over \a}]^N e^{-N}$, and 
we can choose $N=3/\h$; therefore given a couple of fields external to a vertex $v$ with different $\vec x$, we can associate a factor $\g^{2 h_{v'}}$ in the bounds.

On the other hand if there is a $D$ field we get in the bound an extra $\g^{h_{v'}-h_v }$ from
the expression
\be \bar g^{(h_{v'})}(\vec\o\vec x, x_{0,1}-z_0)-\bar g^{(h_{v'})}(\vec\o\vec x,x_{0,2}-z_0)=
(x_{0,1}-x_{0,2}) \int_0^1 dt \partial \bar g^{(h_{v'})}
(\vec\o\vec x,\hat x_{0,1,2}(t)-z_0)
\label{df11} \ee
where $\hat x_{0,1,2}(t)=x_{0,1}+t(x_{0,2}-x_{0,1})$. In conclusion
\begin{enumerate}
\item To each non-resonant $v$ we associate a factor \pref{sap} so that we get in the bound an extra factor
$\prod_{v\in V_\chi}\g^{2 h_v S_v^L}$
\item There is a factor $\prod^*_{v }\g^{h_{ v'}}$ where $v$ 
are the endpoints $\n,\a,\x$ (it
 comes from the definition of $\n$ and the presence $(x_0-y_0)$ or $(\vec\o \vec x-\r\a$).
\item
In the resonant $v$ with $l\ge 2$ fields there is a factor
$
\prod_{v\in H}  \g^{2(h_{v'}-h_v )}$.
For $l=2$ this it is due to the $\RR$ definition,
for $l\ge 4$ by anticommutativity.
\item  In the terms with $|P_v|\ge 8$ we can consider the fields $\psi^\e_x$ whose number is maximal; we can group them in couples connected by path in $\bar T$ non overlapping,
and or have different $\vec x$, hence there is a path in $\bar T$ connecting them giving an extra $\g^{2 h_{v'}}$, or they have the same $\vec x$ so that there is an extra $\g^{2(h_{v'}-h_v)}$. 
This produces an extra $\g^{-\a |P_v|}$, see \S F in \cite{M3}.
\end{enumerate}
We bound first the effective potential ($J=0$).
If $\t\in \TT_{h,n}$, the set of trees with $n$ end-points and defining
\be
||K_{\t,{\bf P},T}^{(h+1)}||={1\over \b L^d}\sum_{\vec x}  \int d x_{0, v_0 } |K_{\t,{\bf P},T}^{(h
+1)}|
\ee
we get
\be
||K_{\t,{\bf P},T}^{(h+1)}||\le  C^n\e_0^n \prod_v {1\over S_v!}\g^{-h_v(S_v-1)}\prod_{v\in V_\chi}\g^{2 h_v S_v^L}
\prod^* _v \g^{h_{ v'}} \prod_{v\in H} 
\g^{2(h_{v'}-h_v )}
 \prod_{v \in V_\chi} \g^{-\a |P_v|}
\label{sssa}
\ee
If the first vertex $v_0\in V_\chi$  is non resonant we get
\be
 \prod_{v\in V_\chi} 
 \g^{-h_v S_v} \prod_v \g^{h_v S^L_v}\prod^* _v \g^{h_{ v'}} \prod_{v\in H, v\not =v_0} 
\g^{h_{v'}}=1
\quad\quad \prod_{v\in V_\chi}  \g^{h_v} \prod_{v\in H, v\not= v_0} 
\g^{-h_v }\le  \g^{h_{v _0}}\label{aaaaa}
\ee
We use that $S_v=S_v^L+S_v^H$,   $\prod_v \g^{h_v S^L_v}= \prod_{v\in L} 
 \g^{h_{v'}} \prod^{**}_{v} 
 \g^{h_{v}} $, with $\prod^{**}_{v} $ is over the first vertex $v \in V_\chi$ after the $\e,\l$ endpoints, and that  $\prod_{v\in L} 
 \g^{h_{v'}}\le \prod_{v\in L} 
 \g^{h_{v'}-h_v}$
\bea
&&||K_{\t,{\bf P},T}^{(h+1)}||\le C^n\e_0^n \g^{h_{v_0} } 
\prod_v {1\over S_v!}\prod_{v\in V_\chi} 
\g^{(h_{v'}-h_v )}  \prod^{**}_{v} 
 \g^{h_v} 
\prod_{v \in V_\chi} \g^{-\a |P_v|}
\label{sssa}
\eea
where $\prod^{**}_{v }$ is over the vertices $v\in V_\chi$ following from the end-points associated to $\e,\l$. 
Note that $\sum_{\bf  P}[\prod_{v\in V_\chi}\g^{-{1\over 8} |P_v|}]\le C^n$; moreover 
$\sum_{\bf T}[\prod_v {1\over S_v!}]\le C^n$.
The sum over the trees $\t$ is done performing the sum of unlabeled
trees and the sum over scales. The unlabeled trees can be bounded by $4^n$ by Caley formula, 
and the sum over the scales reduces to the sum over $h_v$, with $v\in V_\chi$, as
given a tree with such scales assigned, the others are of course determined. 

Let us consider now the case in which the first vertex $v_0$ is resonant; we can distinguish two cases.
If we are considering the contribution to the beta function then there is no $\RR$ applied in $v_0$ so that
the same bound as above is found with $h_{v_0}=h+1$. Instead if $\RR$ is applied we get
instead of \pref{aaaaa}, as there is an extra $\g^{h_{v'_0}
-h_{v_0}}$ 
\be
 \prod_{v\in V_\chi}  \g^{-h_v S_v}\prod_v  \g^{h_v S^L_v}\prod^* _v \g^{h_{ v'}} \prod_{v\in H} 
\g^{h_{v'}}=\g^{h_{v' _0}}
\quad\quad  \prod_{v\in V_\chi}  \g^{h_v} \prod_{v\in H} 
\g^{-h_v }\le  1\label{aaaaa1}
\ee
and the same bound is found, as $h_{v'_0}=h+1$. In conclusion we get
\be
\sum_{\t\in \TT_{h,n}  }\sum_{{\bf P},T} 
 ||K_{\t,{\bf P},T}^{(h+1)}||\le C^n \e_0^n \g^h
\ee
The running coupling constant $\a_h,\x_h$ verify 
\be
\a_{h-1}=\a_h+O(\e_0^2 \g^{h\over 2} )\quad \quad \x_{h-1}=\x_h+O(\e_0^2 \g^{h\over 2} )
\ee
where the factor $\g^{h\over 2}$ is due to the fact that the trees have at least an $\e,\l$ endpoint, from the factor $\prod^{**}_{v} 
 \g^{h_v} $ in \pref{sssa}
 (short memory property).
The flow of $z_h, \a_h$ is therefore summable; in addition
one can choose $\n$ so that $\n_h$ is bounded, by proceeding as in
Lemma 2.7 of cite{M3}.

\section{Decay of correlations}

We consider now the current correlations, which can be written as
\be
H_{\m,\n}(\xx,\yy)
=\sum_{h,n} \sum_{\t\in \TT_{h, n+2}}\sum_{{\bf P},T
} G_{\t,{\bf P},T}(\xx,\yy)
\ee
where $\TT_{h,n+2}$ is the set of trees with $n+2$ end-points, two of them associated to the $J$
end-points.
In the trees $\t$ we can identify a vertex $v_{x}$
for the end-point corresponding to $J_\xx$, and $v_{y}$
for the end-point corresponding to $J_\yy$ with $h_{v_x} =h_{v_y}=+2$; we call $\hat v $, with scale 
$\hat h$, 
the first vertex $v\in V_\chi$ such that $v_x,v_y$ follows $\hat v $, and $v_0$ the first vertex 
$\in V_\chi$, with scale $h$. There are several constraints.
\begin{enumerate}
\item By \pref{fa} and using that $\vec x-\vec y=\sum_{w\in C_{v_x,v_y}  }
 \vec \d_w^{i_w}$ we get 
$n\ge \sum_{w\in C_{v_x,v_y}  } |\vec \d_w^{i_w}|
\ge  |\vec x-\vec y|$
\item $h\ge \bar h(n)$ with, if $|\vec z|=1+\min (|\vec x|,|\vec y|)$
\be
\g^{-\bar h}\le \sup_{\vec q=\sum_{i=1}^n \vec e_i}  {1\over
|| {\vec \o (\vec x+\vec q)-\r \a}||} \le C (|\vec z|+n)^\t
\ee
\end{enumerate}
With respect to the 
bound for the $J=0$ case there are the following differences.
%
If  $T_{\hat v}$ is the tree connecting the 2 $J$ endpoints, we have 
an extra $\g^{\hat h}$ due to the fact that we do not integrate over the coordinates of the $J$
fields, and we can extract from the the propagators in $\prod_{l\in \bar T_{\hat v}} g^{(h_l )}$,
$h_l\ge \hat h$ a decay factor
\be
{1
\over 1+(\g^{\hat h}|x_0-y_0 |)^N}\ee

Moreover there is no $\RR$ in the resonant terms with one or two external $J$ lines.
We can multiply and divide by 
$\g^{-4 \bar h  } \g^{4 \bar h  } $: we can select two paths in $\t$
$v_0< v_1<..v_x$ and $v_0<v'_1<..v_y$, writing
\be
\g^{2 \bar h }=\g^{2(\bar h- h_ {v_1})}...\g^{2 h_ {v'_x}}\quad\quad \g^{2 \bar h }=\g^{2(\bar h- h_ {v'_1})}...\g^{2 h_ {v'_y}}
\ee
where $v'_x$, $v'_y$ are the first vertex $\in V_\chi$ after $v_x$, $v_y$.
We get therefore the following bound
\be
|G_{\t,{\bf P},T}(\xx,\yy)|\le \g^{-4 \bar h  }{C^n|\e|^n \g^{\hat h  }
\over (\g^{\hat h}|x_0-y_0 |)^N} \prod_v {1\over S_v!}\g^{-h_v(S_v-1)}\prod_{v\in V_\chi}\g^{2 h_v S_v^L}
\prod^* _v \g^{h_{ v}} \prod_{v\in H} 
\g^{2(h_{v'}-h_v )}
 \prod_{v \in V_\chi} \g^{-\a |P_v|}
\label{sssa1}
\ee
where $H$ now includes also resonant terms with one or two $J$ fields.
Proceeding as in \S 7 and for $|x_0-y_0|>1$, if $\TT_n$ are the trees with $n$ end-points
\be
\sum_{\t\in\TT_{h,n}} \sum_{{\bf P},T} |G_{\t,{\bf P},T}(\xx,\yy)|\le\g^{-3 \bar h  }{C^n
|\e|^n
\over 1+(\g^{\bar  h}|x_0-y_0 |)^N}\le C^n |\e|^n {|\vec z|^{3 \t}\over (|\vec z|^{-3 \t}
|x_0-y_0|)^N} (1+{n\over |\vec z|})^{(N+3) \t}
\label{sssa2}
\ee
The sum over $h\ge \bar h$ can be bounded by an 
an extra $\g^{-\bar h}$.
As $|\vec z|\ge 1$ and $n/|\vec z|\le n$; we can
sum over $n$ obtaining, remembering the constraint  $n\ge |\vec x-\vec y|$ 
\be
|H_{\m,\n}(\xx,\yy)|\le C {|\vec z|^{4 \t}
\over (|\vec z|^{-3 \t}
|x_0-y_0|)^N} |\e|^{|\vec x-\vec y|/4}
\ee
The analysis of the 2-point function is done in a similar way; there are 2 endpoints associated with the externl fields, so with respect to the bound for the effective potential there is an extra factor $\g^{-2 \bar h }$ and an extra $\g^{\bar h }$ 
from the lack of integration; the sum over the scales produces an extra $|\bar h|$.
\vskip.3cm
{\bf Acknowledgements.} This work has been supported 
by MIUR, PRIN 2017 project MaQuMA, PRIN201719VMAST01.

\bibliographystyle{amsalpha}

\begin{thebibliography}{19}
\bibitem{A0} P. W. Anderson: {\it  Absence of diffusion in certain random lattices.} Phys. Rev. 109, 1492–1505 (1958)
\bibitem{loc0}
J. Froehlich and T. Spencer:{\it 
Absence of diffusion in the Anderson tight binding
model for large disorder or low energy.}
Comm. Math. Phys. 88, 151 (1983)
\bibitem{loc1}
M. Aizenman and S. Molchanov: {\it
Localization at large disorder and at extreme
energies: an elementary derivation.} 
Comm. Math. Phys. 157, 245 (1993)
\bibitem{S} Ya. Sinai: {\it Anderson Localization for one dimensional difference Schroedinger operator with quasiperiodic potential}. 
J. Stat. Phys. 46, 861 (1987)
\bibitem{FS}
J. Froehlich, T. Spencer, T. Wittwer: {\it 
Localization for a class of one-dimensional quasi-periodic Schrödinger operators}.
Comm. Math. Phys.132,1, 5 (1990)
\bibitem{b1}
J. Bourgain.  {\it Anderson localization for quasi-periodic lattice Schroedinger operators on Zd,
d arbitrary. } Geom. Funct. Anal., 17(3):682–706, 2007.
\bibitem{b2}
J. Bourgain, M. Goldstein, and W. Schlag.{\it  Anderson localization for Schroedinger operators on Z
2 with quasi-periodic potential}. Acta Math., 188(1):41–86, 2002
\bibitem{1} Svetlana Jitomirskaya, Wencai Liu, Yunfeng Shi  
{\it Anderson localization for multi-frequency quasi-periodic operators on Zd}
arXiv:1908.03805  
\bibitem{A}
Fleishman, L, and P. W. Anderson (1980), {\it  Interactions and
the Anderson transition,} Phys. Rev. B 21, 2366–2377.
\bibitem{F}
A.M. Finkelstein {\it Influence of coulomb interaction on
the properties of disordered metals}.  Zh. Eksp. Teor. Fiz.
168 (1983)
\bibitem{Gi}
Giamarchi, T, and H. J. Schulz,  {\it Anderson localization and interactions in one-dimensional metals } Phys.
Rev. B 37, 325–340 (1988)
\bibitem{M1a} V.Mastropietro  {\it Small Denominators and Anomalous Behaviour in the Incommensurate Hubbard–Holstein Model}.
Commun. Math. Phys. 201, 81 (1999)
\bibitem{G1a} G. Vidal, D. Mouhanna, T. Giamarchi, 
{\it Correlated Fermions in a One-Dimensional Quasiperiodic Potential} 
 Phys. Rev. Lett. 83, 3908 (1999)
\bibitem{Ba} D.M. Basko, I. Alteiner , B. L. Altshuler: {\it 
Metal-insulator transition in a weakly interacting many-electron system with localized single-particle states.} 
Ann. Phys. 321, 1126 (2006)
\bibitem{H4} A. Pal, D.A. Huse:  {\it 
Many-body localization phase transition. }
Phys. Rev. B 
82,  174411 (2010)
\bibitem{H5}
S. Iyer, V. Oganesyan, G. Refael, D. A. Huse: {\it 
Many-body localization in a quasiperiodic system.}
 Phys. Rev. B 87, 134202 (2013)
\bibitem{H19}
D. A. Abanin, Ehud Altman, Immanuel Bloch, Maksym Serbyn
{\it  Many-body localization, thermalization, and entanglement.}
Rev. Mod. Phys. 91, 021001 (2019)
\bibitem{Z0}
M Schreiber, S. Hodgman, P. Bordia, H. P. Lüschen M H. Fischer, R Vosk, E Altman, U Schneider, Bloch {\it Observation of many-body localization of interacting fermions in a quasirandom optical lattice}
Science 
349, 6250, 842-845 (2015)
\bibitem{Z1}
P. Bordia, H. P. Lüschen, S. S. Hodgman, M. Schreiber, I. Bloch,  and U. Schneider, 
{\it Coupling Identical one-dimensional Many-Body Localized Systems}
Phys. Rev. Lett. 116, 140401 (2016).
\bibitem{a}
V. Khemani, D. N. Sheng, and D. A. Huse, 
{\it Two universality classes for the many-body localization transition}
Phys. Rev.
Lett. 119, 075702 (2017).
\bibitem{b}
P. Naldesi, E. Ercolessi, and T. Roscilde, 
{\it Detecting a many-body mobility edge with quantum quenches}
SciPost Phys.
1, 010 (2016)
\bibitem{c}
F. Setiawan, D.-L. Deng, and J. H. Pixley, 
{\it Transport properties across the many-body localization transition in quasiperiodic and random systems}
Phys. Rev.
B 96, 104205 (2017).
\bibitem{d}
S. Bera, T. Martynec, H. Schomerus, F. HeidrichMeisner, and J. H. Bardarson, 
{\it 
One-particle density matrix characterization of many-body localization}
Annalen der Physik 529,
1600356 (2017)
\bibitem{f}
Y. Wang, H. Hu, S. Chen 
{\it Many-body ground state localization and coexistence of localized and extended states in an interacting quasiperiodic system }
The European Physical Journal B volume 89, 77 (2016) 
\bibitem{H8} M. Znidaric, M. Ljubotina
{\it Interaction instability of localization in quasiperiodic systems}
Proc. Natl. Acad. Sci. U.S.A. 115, 4595-4600 (2018)
\bibitem{K} 
T. Koma, T. Morishita , T.Shuya 
{\it 
Quantization of Conductance in Quasi-Periodic Quantum Wires}
Journal of Statistical Physics volume 174, pages1137–1160 (2019)
\bibitem{g}
A. Purkayastha, S. Sanyal, A. Dhar, and M. Kulkarni
{\it Anomalous transport in the Aubry-André-Harper model in isolated and open systems}
Phys. Rev. B 97, 174206 (2018)
\bibitem{h} T. Cookmeyer, Johannes Motruk, Joel E. Moore
{\it  Critical properties of the many-particle (interacting) Aubry-André model ground-state localization-delocalization transition}
 Phys. Rev. B 101, 174203 (2020)
\bibitem{Ro} V. Ros, M. Mueller, A. Scardicchio	
{\it  Integrals of motion in the Many-Body localized phase}
Nucl. Phys., Section B (2015), 420-465 (2015)
\bibitem{I}J. Z. Imbrie
{\it On Many-Body Localization for Quantum Spin Chains}
Jour. Stat. Phys. 163:998-1048 (2016)
\bibitem{W}
V. Beaud,  S. Warzel 
{\it
Low-Energy Fock-Space Localization for Attractive Hard-Core Particles in Disorder}
Ann. Henri Poincaré 18,3143–3166 (2017)
\bibitem{W1} A. Elgart, A. Klein, G. Stolz, 
{\it Manifestations of Dynamical Localization in the Disordered XXZ Spin Chain}
Comm. Math. Phys., 361, 3, 1083-1113 (2017)
\bibitem{H6}
W. De Roeck, F. Huveneers, 
{\it 
Stability and instability towards delocalization in many-body
localization systems} Phys. Rev. B 95, 155129
(2017).
\bibitem{M1}  V. Mastropietro {\it Localization of interacting fermions in the Aubry-Andr\'e model}
Phys. Rev. Lett. 115, 180401 (2015)
\bibitem{M2} V. Mastropietro:  {\it Localization in the ground state of an interacting quasi-periodic fermionic chain}
Comm. Math. Phys. 342, 1, 217-250 (2016) 
\bibitem{M3} V. Mastropietro
{\it Localization in Interacting Fermionic Chains with Quasi-Random Disorder}
Comm. Math. Phys. 351, 283–309(2017)
\bibitem{G1} G.Gallavotti  {\it Twistless KAM tori}
Comm. in Math. Phys. 164, 145–156 (1994)
\bibitem{Au} S. Aubry {\it  Anti-integrability in dynamical and variational problems}
Physica D 86, 1–2, 1, 284-296 (1995)
\bibitem{Po} V Mastropietro, M Porta
{\it Canonical Drude weight for non-integrable quantum spin chains} 
J. Stat. Phys. 172,  379-397 (2018)
\bibitem{p} B. Bertini, F. Heidrich-Meisner, C. Karrasch, T. Prosen, R. Steinigeweg, M. Znidaric
{\it  Finite-temperature transport in one-dimensional quantum lattice models} 
Rev. Mod. Phys. (2020) 
\bibitem{La} R de la Llave Tutorial on KAM theory, American Mathematical Society, 2003
\bibitem{M4a} V. Mastropietro
{\it Conductivity in the Heisenberg chain with next-to-nearest-neighbor interaction} 
Phys. Rev. E 87, 042121 (2013)
\bibitem{M4} V. Mastropietro {\it
Interacting spinning fermions with quasi‐random disorder}
Annalen der Physik 529, 7
1600270 (2017)
\bibitem{M5} V. Mastropietro 
{\it Dense gaps and scaling relations in the interacting Aubry-Andre' model}
Phys. Rev. B 95, 075155  (2017)
\bibitem{Z}
P. Prelovsek, O.S. Barisic, M. Znidaric 
{\it
Absence of
full many body localization in disordered Hubbard chain}
Phys. Rev. B 94, 241104 (2016)
\bibitem{M6} V. Mastropietro
{\it Coupled identical localized fermionic chains with quasi-random disorder}
Phys. Rev. B 93, 245154 (2016)
\bibitem{M7} V. Mastropietro  {\it Persistence of gaps in the interacting anisotropic Hofstadter model}
Phys. Rev. B 99, 155154  2019
\bibitem{M8} V. Mastropietro 
{\it Stability of Weyl semimetals with quasiperiodic disorder} 
Phys. Rev. B 102, 04510 2020
\end{thebibliography}

\end{document}